\documentclass[twocolumn,showpacs,preprintnumbers,amsmath,amssymb]{revtex4}

\usepackage{graphicx}
\usepackage{dcolumn}
\usepackage{bm}

\begin{document}

\title{Vortex-flow electromagnetic emission in stacked intrinsic Josephson junctions}


\author{Myung-Ho Bae}
\author{ Hu-Jong Lee}%

\affiliation{Department of Physics, Pohang University of Science
and Technology, Pohang 790-784, Republic of Korea}%

\date{\today}

\begin{abstract}
We confirmed the existence of the collective transverse plasma
modes excited by the motion of the Josephson vortex lattice in
stacked intrinsic Josephson junctions of
Bi$_2$Sr$_2$CaCu$_2$O$_{8+x}$ by observing the multiple
subbranches in the Josephson-vortex-flow current-voltage
characteristics. We also observed the symptom of the microwave
emission from the resonance between the Josephson vortex lattice
and the collective transverse plasma modes, which provides the
possibility of developing Josephson-vortex-flow electromagnetic
oscillators.
\end{abstract}

\pacs{74.72.Hs, 74.50.+r, 74.78.Fk, 85.25.Cp }
\maketitle

The generation of THz-range electromagnetic waves using the
Josephson-vortex dynamics in naturally stacked
Bi$_2$Sr$_2$CaCu$_2$O$_{8+x}$ (Bi-2212) intrinsic Josephson
junctions (IJJs) has been attempted extensively because of the
coherent (thus, high-power), continuous, and frequency-tunable
characters of the generated waves \cite{Tachiki}. The characters
of this technique are distinctive from those by other attempts
\cite{THz} based on quantum cascade, relativistic electron
bunches, and optical parametric control. This electromagnetic
emission from the Josephson vortex system is induced by the
resonance between moving Josephson vortices and the collective
transverse plasma (CTP) modes in the stacked IJJs
\cite{Inductive}.

A stack of Bi-2212 containing $N$ IJJs exhibits $N$ eigen CTP
modes, which can be excited by the moving Josephson vortex lattice
(JVL) that forms in a high magnetic field \cite{Machida}. If the
frequency of the temporal oscillation of the phase difference
across a junction due to moving JVL matches with that of a
transverse plasma mode, a resonant plasma oscillation is excited
with microwave emission at the boundary of stacked IJJs. The JVL
also transforms its lattice configuration along the $c$-axis
direction in accordance with the $c$-axis standing-wave mode of
the strongly amplified plasma oscillations. The resonance of JVL
to the CTP modes appears as the multiple collective Josephson
vortex-flow branches in the tunneling current-voltage ({\it I-V})
curves of stacked junctions \cite{Bae1}.

We report the observation of the CTP modes induced by the JVL
motion and the excitation of corresponding electromagnetic waves
in a stack of IJJs in Bi-2212. The existence of the CTP modes was
confirmed by observing the multiple branches in the Josephson
vortex-flow region. In addition, for a proper bias, the emission
of the electromagnetic waves by the collective vortex resonance
motion in a stack of IJJs (the oscillator stack) was examined
using another stack of IJJs (the detector stack), which was placed
within a fraction of $\mu$m from the oscillator stack. The
microwave emission from the oscillator stack and the resulting
irradiation onto the detector stack was evidenced by the
suppression of the tunneling critical current revealed in the
quasiparticle branches and the increase of Josephson vortex-flow
voltages in the detector stack. Our numerical calculation for the
effect of microwave irradiation on the detector stack was also
consistent with our observed results.

\begin{figure}[t]
\begin{center}
\leavevmode
\includegraphics[width=0.7\linewidth]{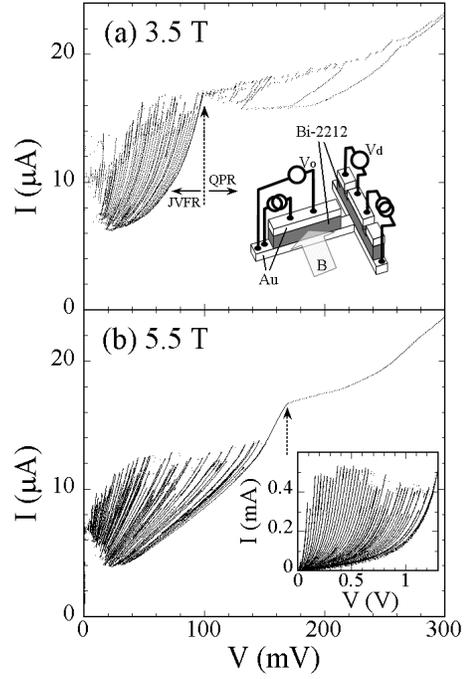}
\caption{The $H$-field dependence of the Josephson-vortex-flow
branches and the quasiparticle branches of SP1 in (a) $H$=3.5 T
and (b) 5.5 T. Inset of (a): the sample configuration. Inset of
(b): multiple quasiparticle branches of SP1 in zero field.}
\end{center}
\end{figure}

Bi-2212 single crystals, prepared by the conventional
solid-state-reaction method, were slightly overdoped. We
fabricated, using the double-side cleaving technique \cite{Wang},
two samples of IJJs, each sandwiched between two Au electrodes at
its top and bottom without the basal part [the inset of Fig.
1(a)]. Adopting the geometry with the basal part eliminated
enabled us to measure the Josephson vortex dynamics in coupled
IJJs without the interference of the vortex motion in the basal
stack. The detailed fabrication procedure is described in Ref.
\cite{Bae2}. In the inset of Fig. 1(a) the left and right stacks
are the oscillator and detector stacks, respectively.  The lateral
size of each oscillator of two samples was 16 $\times$ 1.5
$\mu$m$^2$ [SP1] and 15 $\times$ 1.4 $\mu$m$^2$ [SP2],
respectively. The magnetic field was aligned in parallel with the
junction planes within 0.01 degree to avoid the pinning of
Josephson vortices by the formation of pancake vortices in CuO$_2$
bilayers.

Figs. 1(a) and (b) show the magnetic field dependence of {\it I-V}
curves of SP1 at 4.2 K in 3.5 and 5.5 T, respectively. The contact
resistance caused by the two-terminal configuration adopted was
subtracted numerically. The current-bias point corresponding to
the dotted vertical arrow in each figure originates from the
zero-field Josephson critical current. Thus, the voltage-bias
region below the critical point corresponds to the pair-tunneling
state, although it is resistive due to the Josephson vortex flow
in high fields. The set of multiple branches in the voltage
(McCumber) bias region above the critical point are the
quasiparticle branches, while those below the critical point are
the collective Josephson vortex-flow branches \cite{Bae1}.
Estimated from the number of zero-field quasiparticle branches as
in the inset of Fig. 1(b) for SP1, stacks in SP1 and SP2 contained
45 and 22 IJJs, respectively.

The numbers of the Josephson-vortex-flow branches below the
critical bias points for SP1 and SP2 are $\sim$42 and $\sim$18
(not shown), respectively, which are similar to the numbers of the
IJJs in the respective samples. In addition, the multiple
Josephson-vortex-flow branches become clearer and wider for a
higher transverse magnetic field, which is in contrast to the
shrinking quasiparticle branches with increasing fields [see Fig.
1(a) and (b)].

The collective transverse plasma oscillation excited by the moving
JVL is expected to emit electromagnetic waves at the junction edge
\cite{Hech2}. Detection of this microwave emission, in turn, would
be more positive confirmation that the observed multiple branches
resulted from the moving JVL in resonance with the CTPs. A
separate stack for the microwave detection was positioned a
fraction of $\mu$m apart from the oscillator stack. In the
direction facing the applied field, the detector was 0.7 $\mu$m
wide, which was longer than the Josephson penetration depth of 0.3
$\mu$m and thus was in a long-junction limit. The oscillator and
the detector stacks, connected by the bottom Au common-ground
electrode, acted as a microwave coupler [see the inset of Fig.
1(a)]. The frequency of emitted microwaves may reach a THz range
so that a conventional Nb-based Josephson detector cannot be used,
because its gap size is smaller than the energy of the emitted
waves.

\begin{figure}[t]
\begin{center}
\leavevmode
\includegraphics[width=0.7\linewidth]{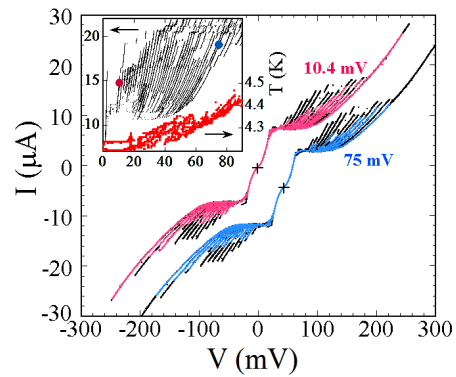}
\caption{(Color online) Response of the \emph{quasiparticle
branches} of the detector stack of SP1 to biasing the oscillator
stack. The curves are shifted for clarity. Inset: the bias
condition of the oscillator stack as denoted by dots and the
corresponding local temperature of the detector stack, with the
base temperature at 4.2 K, for varying bias in the oscillator
stack.}
 \end{center}
\end{figure}

The response of IJJs in the detector stack to the emitted
microwaves from the oscillator stack may be revealed as the
appearance of the Shapiro steps, the suppression of the $c$-axis
tunneling critical current, $I_c$, revealed in the quasiparticle
branches, or the increase of the voltage of the low-bias region
due to the resistive motion of microwave-induced vortices
\cite{Micro,Doh}. The inset of Fig. 2 shows the bias conditions,
$V_{osc}$, displayed by dots in the Josephson vortex-flow branches
of the oscillator stack of SP1 in 3 T. The black {\it I-V} curves
in the main panels are the \emph{quasiparticle branches} of the
detector stack without any bias in the oscillator stack. For the
biases of $V_{osc}$=10.4 and 75 mV (corresponding to 110 and 800
GHz \cite{conversion}, respectively) in the oscillator, $I_c$ in
the \emph{quasiparticle branches} of the detector stack is
suppressed significantly. Our previous studies \cite{Bae2, Doh}
confirmed that the irradiation of microwaves on the IJJs
suppresses the $I_c$ revealed in the quasiparticle branches.

The temperature increase of the detector stack by the self-heating
in the oscillator for a finite dc bias may cause the similar
suppression of the $I_c$ \cite{Kume}. We directly monitored the
actual temperature variation in the detector for any finite bias
in the oscillator. Details of the thermometry are explained in
Ref. \cite{Bae3}, which confirms that both stacks in our
measurement configuration should be at an identical temperature.
The inset of Fig. 2 shows that the actual temperature increase in
this measurement turned out to be less than 4.5 K, so that the
observed suppression of the $I_c$ around 4.2 K should not have
been caused by the bias-induced self-heating. One may also
attribute the behavior to the leakage of the bias current from the
oscillator to the detector through the bottom Au common-ground
electrode, rather than the microwave emission. However, any shift
of the {\it I-V} curves due to a dc offset current in the detector
was not observed for any biases used. We thus attribute the
suppression of the $I_c$ in the quasiparticle branches to the
response of the stacked IJJs in the detector to the mircowave
irradiation.

\begin{figure}[t]
\begin{center}
\leavevmode
\includegraphics[width=0.7\linewidth]{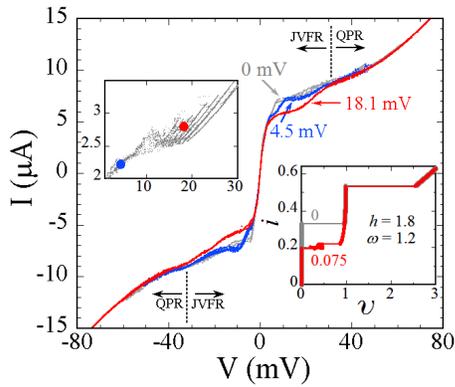}
\caption{(Color online) The change in the \emph{Josephson
vortex-flow branch} in the detector of SP2 for the bias voltage of
its oscillator $V_{osc}$=0, 4.5, and 18.1 mV. Upper inset: bias
condition (denoted by dots) of the oscillator stack. Lower inset:
numerically calculated {\it I-V} curves for the ac ($\omega$=1.2)
field component of $h_{ac}$=0 and 0.075 in a $dc$ magnetic field
of $h$=1.8.}
\end{center}
\end{figure}

Fig. 3 shows the change in the \emph{Josephson vortex-flow branch}
in the detector of the specimen SP2 with varying the bias voltage
$V_{osc}$ of its oscillator. For $V_{osc}$=4.5 mV (corresponding
to 100 GHz) little variation of the {\it I-V} curves from the
zero-bias condition is visible. The bias of 4.5 mV corresponds to
the near-triangular JVL in the oscillator stack, which represents
the out-of-phase Josephson vortex distribution between adjacent
junctions and thus corresponds to the weakest emission power, as
illustrated in Figure 3. For $V_{osc}$=18.1 mV (corresponding to
400 GHz), however, a distinct increase of the Josephson
vortex-flow voltage from the zero-bias condition is seen. We
interpret it as arising from the flow of additional Josephson
vortices generated by the magnetic-field component of emitted
higher-intensity microwaves from the oscillator with JVL close to
the highly coherent square-lattice configuration.

We numerically calculated vortex-flow feature, in the detector
stack induced, by the magnetic-field component of any irradiated
microwaves \cite{Goldobin}. For simplicity, we consider a detector
consisting of a single Josephson junction. The perturbed
sine-Gordon equation for the dynamics of the phase difference
across a Josephson junction, $\phi$, is
\begin{equation}
\frac{\partial^2 \phi}{\partial x^2}-\frac{\partial^2
\phi}{\partial t^2}-\mbox{sin}\phi = \alpha \frac{\partial
\phi}{\partial t}-i-i_{ac}\mbox{sin}\omega t,
\end{equation}
where the spatial and time variables of the $\phi$ are normalized
by the Josephson penetration depth $\lambda_J$ and by the inverse
of the Josephson plasma frequency, $\omega_p^{-1}$, respectively.
The coefficient $\alpha$ is the damping parameter due to the
quasiparticle tunneling, $\omega$ is the ac frequency normalized
by $\omega_p^{-1}$ and $i$ ($i_{ac}$) is the $dc$ ($ac$) bias
current normalized by the tunneling critical current $I_c$. We
adopted the boundary condition in the external $dc$ and $ac$
magnetic fields as $\partial \phi/\partial
x(x=0,t)=h-h_{ac}\mbox{sin}\omega t$ and $\partial \phi/\partial
x(x=l,t)=h$, where $l$ is the junction length normalized by
$\lambda_J$, $h\equiv2\pi N_f/l$, and $N_f$ is the number of the
Josephson vortices. The averaged voltage, {\it v}, in the $x$ axis
is normalized as $\langle
\partial \phi/\partial t\rangle l/2\pi$.

The lower inset of Fig. 3 shows the simulation results for a $dc$
magnetic field of $h$=1.8 without and with the irradiation of the
external electromagnetic waves, where we used parameter values of
$l$=3, $\omega$=1.2, and $\alpha$=0.1. The simulation for the
microwave irradiation mimics the observation in the detector as in
Fig. 3: the appearance of the finite Josephson-vortex-flow voltage
for 0.2$<i<$0.32 in the inset. The vertical step at $v$=1 is the
resonant Josephson-vortex-flow branch. The numerical results
suggest that the Josephson vortices induced by the $ac$
magnetic-field component of the electromagnetic waves effectively
play the role of a $dc$ magnetic field of the corresponding
magnitude, which is qualitatively consistent with our experimental
observation in Fig. 3 and previous reports \cite{Barkov}.

The microwave emission from the resonance between the Josephson
vortex motion and the collective plasma oscillation modes, as
manifested by changing {\it I-V} curves in the detector stack,
provides a possibility of application to the Josephson vortex-flow
oscillator. The observation of the Shapiro steps in the detector
stack would be the most convincing evidence for the microwave
generation from the oscillator stack. Recently it was
theoretically predicted \cite{Tachiki} that the radiation power
from Bi-2212 Josephson-vortex-flow oscillator in an optimal
condition can be as high as $\sim$2000 W/cm$^2$ from the edge of
the stack, which corresponds to about 1$\sim$2 $\mu$W in our
samples. Observing Shapiro steps requires sufficiently strong
emission power with improved matching between the oscillator and
detector stacks. Thus, devising optimal impedance matching
schemes, using antennas or dielectric wave guides \cite{Tachiki},
is the most essential step for the decisive confirmation of the
microwave emission as well as the practical application of the
Josephson vortex-flow motion to the electromagnetic local
oscillator.

This work was supported by the National Research Laboratory
program administrated by Korea Science and Engineering Foundation
and also by the AFOSR/AOARD of the US Air Force under Contract No.
FA5209-04-P-0253.


\begin{thebibliography}{00}
\bibitem{conversion} The frequency is estimated based on the frequency-voltage (per junction)
                     conversion relation, $f$=$2eV_{osc}/Nh$ (=483.6 GHz/mV$\times V_{osc}/N$).
\end{thebibliography}
\end{document}